\documentclass[12pt]{article}

\begin{document}

\begin{titlepage}

\begin{flushright}
IUHET-497\\
hep-ph/0608094
\end{flushright}
\vskip 2.5cm

\begin{center}
{\Large \bf Limits on Neutron Lorentz Violation \\
From Pulsar Timing}
\end{center}

\vspace{1ex}

\begin{center}
{\large Brett Altschul\footnote{{\tt baltschu@indiana.edu}}}

\vspace{5mm}
{\sl Department of Physics} \\
{\sl Indiana University} \\
{\sl Bloomington, IN 47405 USA} \\

\end{center}

\vspace{2.5ex}

\medskip

\centerline {\bf Abstract}

\bigskip

Pulsars are the most accurate naturally occurring clocks, and data about them can be
used to set bounds on neutron-sector Lorentz violations. If $SO(3)$ rotation symmetry
is completely broken for neutrons, then pulsars' rotation speeds will vary
periodically. Pulsar timing data limits the relevant Lorentz-violating coefficients
to be smaller than $1.7\times10^{-8}$ at at least 90\% confidence.

\bigskip

\end{titlepage}

\newpage

Since the discovery that Lorentz symmetry could be
broken spontaneously in string theory~\cite{ref-kost18}, there has been a great deal
of interest in the study of Lorentz violation. Besides string theory, many
other candidate theories of quantum gravity also predict Lorentz
violation, at least in certain regimes. If Lorentz violations were observed
experimentally, they would be of tremendous importance, potentially telling us a
great deal about Planck scale physics. Experimental searches for Lorentz violation
have thus far failed to produce any positive results.
These searches have included studies of matter-antimatter asymmetries for
trapped charged particles~\cite{ref-bluhm1,ref-bluhm2,ref-gabirelse,
ref-dehmelt1} and bound state systems~\cite{ref-bluhm3,ref-phillips},
determinations of muon properties~\cite{ref-kost8,ref-hughes}, analyses of
the behavior of spin-polarized matter~\cite{ref-kost9,ref-heckel2},
frequency standard comparisons~\cite{ref-berglund,ref-kost6,ref-bear,ref-wolf},
Michelson-Morley experiments with cryogenic resonators~\cite{ref-muller1,
ref-stanwix}, Doppler effect measurements~\cite{ref-saathoff,ref-lane1},
measurements of neutral meson
oscillations~\cite{ref-kost10,ref-kost7,ref-hsiung,ref-abe},
polarization measurements on the light from distant galaxies~\cite{ref-carroll1,
ref-carroll2,ref-kost11,ref-kost21}, analyses of the radiation emitted by energetic
astrophysical sources~\cite{ref-jacobson1,ref-altschul6}, and others.

Although the work by Kosteleck\'{y} and Samuel on string theory~\cite{ref-kost18} has
been a
major impetus for the study of Lorentz violation in the last ten years, there are
also other situations in which Lorentz violation is a possibility.
While there is no experimental evidence for Lorentz violation yet, there are numerous
theoretical reasons to believe that Lorentz violation may be possible. A Lorentz- and
CPT-violating effective field theory, the standard model extension (SME), has been
developed in detail~\cite{ref-kost1,ref-kost2,ref-kost12},
to give a general effective field theory parameterization of all possible forms of
Lorentz violation.  The SME framework is useful because it does not tie us to any
particular underlying mechanism for Lorentz symmetry breaking.
Many basic issues in the SME, including stability
and causality~\cite{ref-kost3} and one-loop renormalization~\cite{ref-kost4,
ref-altschul5}, have been examined. The SME contains coefficients parameterizing all
possible Lorentz violations. The experiments mentioned above have placed very tight
constraints on some of these coefficients; however, the bounds of many of the SME
coefficients are still somewhat muddled, and some of the coefficients have not been
bounded at all.

Some of the most accurate tests of Lorentz invariance are clock comparison
experiments. Today's atomic clocks are extremely accurate, and this makes them
wonderful tools for making precision measurements. On the other hand,
the most accurate naturally
occurring clocks are pulsars, and it is natural to ask whether observations of
these rotating neutron stars could also be used to test Lorentz symmetry in a useful
fashion. If fact, we shall see that pulsar data can be used to place some interesting
bounds on Lorentz violations in the neutron sector.

A certain subset of the SME coefficients will be relevant for our analysis of pulsar
data. The Lagrange density for the neutrons in our theory is
\begin{equation}
\label{eq-L}
{\cal L}=\bar{\psi}[(\gamma^{\mu}+c^{\nu\mu}\gamma_{\nu})
i\partial_{\mu}-m]\psi,
\end{equation}
where $\psi$ is the neutron field.
The use of covariant notation is possible because the theory is invariant under
observer Lorentz transformations. That is, the physics is independent of the choice
of coordinates used by an observer. The theory is not invariant under particle
Lorentz transformations, however.

$c$ contains nine parameters that contribute to Lorentz-violating physics at leading
order. The trace $c^{\mu}\,_{\mu}$ only affects the overall normalization of the
electron
field, and the antisymmetric part of $c$ has no effects at first order, where it
is equivalent to a redefinition of the Dirac matrices. There are many other possible
Lorentz-violating terms in the neutron sector. However, the $c$ terms
should be the most important in the context of pulsar timing. Although we observe
pulsars through the electromagnetic radiation they emit, the details of these
radiation processes are unimportant, and we do not need to consider possible Lorentz
violations in the photon sector. The neutrons are of course also subject to
gravitational interactions, but we shall assume that the gravity sector is
conventional. Angular momentum is not conserved according to (\ref{eq-L}), and this
is the effect that we shall seek to exploit; this is similar to what was done
in~\cite{ref-nordtvedt}, where the precession of the sun's axis
in a Lorentz-violating background was considered.

The angular momentum nonconservation effects considered here are quite different
than those studied in the torsion pedulum experiment~\cite{ref-heckel2},
which looks for Lorentz-violating effects in a spin system. The spin pedulum
experiment is only sensitive to axial vector effects, because the spin operator is
itself an axial vector. However, it
should not be surprising that a given experiment will
only be sensitive to a subset of the possible Lorentz-violating coefficients.
As we shall see, pulsar timing experiments are most sensitive to the $c$ term, which
does not have an axial vector form.

The best current bounds on the neutron $c$ coefficients come from reanalysis of
older clock comparison experiments~\cite{ref-kost6}. However, most of the bounds
combine the neutron $c$ with other coefficients for Lorentz violation, and there
could potentially be cancellations between the different coefficients. Only one of
the four experiments analyzed in~\cite{ref-kost6} provides bounds on the
neutron $c$
coefficients only, and the resulting bounds only cover two of the five possible
coefficients that are observable nonrelativistically. None of the clock experiments
provide bounds on the combination $c_{Q}=c_{XX}+c_{YY}-2c_{ZZ}$, where $X$, $Y$, and
$Z$ are coordinates in a sun-centered celestial equatorial frame.
Moreover, any results gleaned
from reanalysis of old experimental data must be considered very cautiously. It is
always possible that the experimenters made some unrecorded assumptions when they
constructed the experiment, which could throw the results, when analyzed in the
context of Lorentz violation, into doubt.

So it would be nice to find alternative bounds on neutron-sector
Lorentz violations, and pulsars are natural systems to consider.
Our model of a pulsar will be rather crude, but it should allow us to pick out
the dominant Lorentz-violating effects. The most important assumptions that we shall
make are that a pulsar is indeed composed of predominantly neutrons and that it is
supported by degeneracy pressure. For more exotic
objects like quark stars, similar bounds ought to be available, but determining them
would be substantially more complicated.

The strength of the bounds we may derive will be determined by the
empirically observed
regularity of the pulses we observe from neutron stars. Real pulsars will wobble
slightly even in the absence of Lorentz violation, because they are not perfectly
rigid rotating systems. The magnetic and spin axes of a pulsar are not aligned,
which automatically implies some asymmetry in the internal angular momentum
distribution.
If conventional mechanisms of wobbling produce larger effects than
any Lorentz violations, then no Lorentz-violating effects would be observable, and
this limits the accuracy of our bounds on $c$.
Moreover, the ultimate limit on the accuracy of our bounds is not
necessarily the stability of the pulsars
themselves.  Effects in a
pulsar's environment may introduce timing errors as well. Even if the origins of
these timing errors are fairly well understood, the quantitative details of how the
observed pulses are affected must be determined from empirical fits to the data.
A Lorentz violation signal could be hidden amidst the effects of more conventional
physics; we must therefore consider the case in which Lorentz violation is the
largest source of
precession and timing errors in the system.
This simplifies the analysis, but it limits the accuracy of our bounds. 

All the fermion coefficients in the minimal SME, except $c$ and one unphysical
coefficient $a$, have
nontrivial Dirac matrix structures. The sign of any contributions from the other
coefficients will depend on the spin states of particles. However, the neutrons in
a neutron star are not in a coherent spin state. This results in averaging
that will diminish the effects of any spin-dependent Lorentz-violating operators.
So only the effects associated with the $c$ coefficients
will receive coherent contributions from all the neutrons in a pulsar.
We are
therefore justified in considering only the $c$ coefficients in the neutron sector.

We shall also consider only those components of $c$ that have observable effects
on nonrelativistic physics.
Although the neutrons in a pulsar are moving, this motion is not highly relativistic.
For a pulsar with a very short rotation period of 2 ms
and a radius of 10 km, the surface speed is $v\approx0.1$. Any
effects caused by $c_{0j}$ or $c_{00}=-c_{jj}$ terms will be suppressed by at least
one factor of $v$.
In reality, the suppression will be substantially greater,
because most of the neutrons in
the pulsar are located closer to the core and are moving more slowly
than those at the surface; also, there
will be some cancellation among the relativistic effects, because neutrons
on opposite sides of the rotating pulsar are moving in opposite directions. In light
of these facts, and because we shall be considering pulsars with periods
significantly larger than 2 ms, we shall neglect all Lorentz-violating effects
except those due to $c_{jk}$.

The suppression of the observable effects of $c_{0j}$ and $c_{00}$ by powers of $v$
deserves some explanation. If we look at the energy-momentum relation, the absolute
magnitude of the contributions proportional $c_{0j}$ or $c_{00}$ are not suppressed.
Nonrelativistically, in fact, a contribution involving, say, $c_{01}$ will be larger
than one involving $c_{12}$ by a factor of ${\cal O}(m/|p_{2}|)$. However, no
contribution can be identified as a signature of Lorentz violation unless its
effects on particles with different rest coordinates are compared. $c_{0j}$ and
$c_{00}$ break boost invariance, and so have different effects for particles in
different rest frames. These different rest
frames are connected by Lorentz boosts. The observable differences in
quantities come from the $-\vec{v}\cdot\vec{p}$ part of the boost transformation
expression $\gamma(E-\vec{v}\cdot\vec{p})$ and so are suppressed by one factor of
${\cal O}(\vec{v}\cdot\vec{p}/m)$ for each time index on $c_{\mu\nu}$. The net
result is a suppression of observable boost invariance violation effects relative to
rotation invariance invariance ones by powers of $v$. This general argument has been
verified by the calculation of specific quantities in, for example,~\cite{ref-kost6}.

The Lorentz-violating $c_{jk}$ coefficients will affect the moments of inertia of a
neutron star. These modifications are easy to determine, because to a very good
approximation, a neutron star is spherical.
We are justified in neglecting any deviations from sphericity not related to Lorentz
violation by our assumption that the largest wobble effects that we
observe are due to the Lorentz violation.
The holds even with the Lorentz violation, because the determination of the star's
shape is basically a static problem, and $c$ does not affect the
isotropy of the neutrons' degeneracy pressure~\cite{ref-colladay}.

To leading order, the effect of the $c_{jk}$ is simply to replace the standard
kinetic energy, $\vec{p}\,^{2}/2m$, with  $(\vec{p}\,^{2}-2c_{jk}p_{j}
p_{k})/2m$, and we see that only the symmetric part $c_{(jk)}=c_{jk}+c_{kj}$
contributes. In some coordinate frame, $c_{(jk)}$ will be diagonal, and this is
the frame of the principal axes. To calculate the effective
moment of inertia around the $z$-axis in this frame, we may imagine that the
star is rotating around this axis. The kinetic energy
(and hence $I_{33}$) is then proportional by an integral
over the mass distribution, with
the integrand proportional to $(1-2c_{11})p_{1}^{2}+(1-2c_{22})p_{2}^{2}$
(instead of simply
$p_{1}^{2}+p_{2}^{2}$). Since the mass distribution is symmetric around the axis of
rotation, the $p_{1}^{2}$ and $p_{2}^{2}$ terms contribute equally, and the net
effect is to multiply the conventional expression for the moment of inertia $I_{33}$
by $1-\left(c_{11}+c_{22}\right)$. Since the overall normalization of the
moment of inertia is an empirically determined quantity, any Lorentz-violating
effect that contributes only to the trace $I_{jj}$ will not be detectable, and an
overall rescaling of $I$ allows us to replace the factor $1-c_{11}-c_{22}$ with
$1+c_{33}$. Therefore
the observable effect of Lorentz violation will simply be to modify the moment of
inertia to $I_{0}\left(1+c_{33}\right)$, where $I_{0}$ is its value in
the absence of rotation invariance violation. It immediately follows that the
complete tensor of inertia is $I_{jk}=I_{0}\left[\delta_{jk}+\frac{1}{2}c_{(jk)}
\right]$.

The preceding arguments suppose that the distribution of the neutrons in momentum
space is not significantly modified by the Lorentz violation; however, this is a
well justified supposition. A neutron star can be
roughly seen as a sphere of degenerate
neutron matter confined by a gravitational field. The structure of the Fermi
sea is largely unchanged by the Lorentz violation. Most occupied neutron states
in a star lie deep in the Fermi sea, and these states will be occupied regardless of
whether or not Lorentz violation exists; this is true whether the Lorentz violation in
question is in the neutron sector or the strong interaction sector. The shape of the
Fermi surface will be
distorted, but this only affects the occupations numbers of states near the
top of the sea. This is a small fraction of the total number of
neutrons present, so any resulting effects will be suppressed; unlike the
changes to the moments of inertia, these are not coherent effects in which all the
particles in the Fermi sea contribute equally.

So the central question is how the modified $I$ affects the observed pattern of
pulses. In fact, a nonzero $c$ will generally lead to slow periodic modulations in
the pulse frequency as the axis of rotation wobbles. Working out this effect can
be tricky; however, we are aided by our empirical knowledge that the rotation is
extremely stable. Therefore, we are justified in assuming that any wobbles are small
oscillations around a stable rotation axis, which according to the ``tennis racket
theorem'' must be the principal axis with either the
largest or smallest moment of inertia. If the
dominant effect causing the spin precession is Lorentz violation,
we may separate $\vec{\omega}$ into $\vec{\omega}_{0}+
\vec{\delta\omega}$, where
$\vec{\omega}_{0}$ lies along one of the eigenvectors of $c_{(jk)}$, and
$\vec{\delta\omega}$ is the fluctuating part.

We shall consider
$\vec{\delta\omega}$ as a small correction, but we do not
really expect it to be all that
much smaller than $\vec{\omega}_{0}$. The spin orientation of pulsar is
determined by the
orientation of its stellar progenitor, which is essentially random. The
quantity $\left\langle\left|\vec{\delta\omega}\right|\right\rangle
/\left|\vec{\omega}_{0}\right|$, with the
time average of the fluctuating part, 
characterizes just how closely aligned $\vec{\omega}$
is with an eigenvector of $c_{(jk)}$; for a particular neutron star, the probability
that this quantity is smaller than, for example,
$\frac{1}{3}$ is less than $4\left[\pi\left(\frac{1}{3}\right)^{2}\right]/4\pi\sim
10^{-1}$. (The factor of four comes from the fact that there are
four stable directions along which $\vec{\omega}_{0}$ may lie.) However, while
we do not expect $\vec{\delta\omega}$ to be strongly suppressed, the leading order
analysis
still captures the essential behavior; including possible higher-order corrections
does not weaken the calculated bounds on $c$ in any way.

In the principal axis frame, the Euler equations are, to leading order in
$\vec{\delta\omega}$,
\begin{equation}
\label{eq-matrix}
\left[
\begin{array}{c}
\dot{\delta\omega}_{1} \\
\dot{\delta\omega}_{2} \\
\dot{\delta\omega}_{3}
\end{array}
\right]=\left[
\begin{array}{ccc}
0 & (c_{22}-c_{33})\omega_{03} & (c_{22}-c_{33})\omega_{02} \\
(c_{33}-c_{11})\omega_{03} & 0 & (c_{33}-c_{11})\omega_{01} \\
(c_{11}-c_{22})\omega_{02} & (c_{11}-c_{22})\omega_{01} & 0
\end{array}
\right]\left[
\begin{array}{c}
\delta\omega_{1} \\
\delta\omega_{2} \\
\delta\omega_{3}
\end{array}
\right].
\end{equation}
It is obvious from the form of these equations that the physical effects can
depend only on the differences in the eigenvalues of $c_{(jk)}$.
Since $\vec{\omega}_{0}$ is along one of the principal axes, two of its components
vanish. Taking the nonzero component to
be $\omega_{03}$, the eigenvalues of the matrix in (\ref{eq-matrix}) are
approximately $0$ and $\pm i\omega_{03}\sqrt{(c_{33}-c_{11})(c_{33}-c_{22})}$.
So the instantaneous angular
velocity $\vec{\omega}$ will undergo oscillations with frequency
\begin{equation}
\Omega=\omega_{0}\sqrt{(c_{33}-c_{11})(c_{33}-c_{22})}.
\end{equation}

$\Omega^{-1}$ sets the scale of the time variation in the rotation frequency
(unless $c_{11}$ and $c_{22}$ are equal); our bounds come from the fact that the
$\Omega$ arising from Lorentz violation
must be smaller than the frequency of the fastest observed modulation in
$\omega=\left|\vec{\omega}\right|$.
The instantaneous
rotation frequency is a quantity that we can observe directly, because the pulsar
beam sweeps over us every rotation. 
%
The magnitude of the angular velocity varies with time as
\begin{equation}
\left|\vec{\omega}\right|\approx\left|\vec{\omega}_{0}\right|+
\frac{\left\langle\left|\vec{\delta\omega}\right|\right\rangle^{2}}
{4\left|\vec{\omega}_{0}\right|}(c_{11}-c_{22})\cos(2\Omega t+\psi),
\end{equation}
where we have evaluated each of the terms to lowest nonvanishing order in $c$.
If the angular momentum is oriented precisely along an eigenvector of $c_{(jk)}$,
there is no variation in $\left|\vec{\omega}\right|$ and hence no observable
effect. However, we have already argued that a very precise alignment of
$\vec{\omega}$ and an eigenvector is unlikely, and that it is almost 90\% likely
that $\alpha\equiv\left\langle\left|\vec{\delta\omega}\right|\right\rangle
/\left|\vec{\omega}_{0}\right|>\frac{1}{3}$.

There will be no time variation in 
$\omega$ if any two of the eigenvalues of $c_{(jk)}$
are equal.
This complicates our task of setting bounds on the coefficients.
If $SO(3)$ rotation invariance is broken, yet there is still an unbroken $SO(2)$
subgroup of the Lorentz group (so that the universe has cylindrical symmetry),
we would expect to see no signal. The $c_{Q}$ part
of $c$ has cylindrical symmetry, so it cannot be bounded by these
measurements.

The bounds we can derive will
therefore be somewhat curious in form. If the lowest frequency of
any observed modulations of $\omega$ is $\varpi$, and the instantaneous angular
frequency is known to an accuracy of $\Delta\omega$, then {\em either}
$2\Omega<\varpi$ {\em or} $|c_{11}-c_{22}|<4\alpha^{-2}\Delta\omega/\omega_{0}$. The
second possibility occurs if the pulsar's wobbles result in a change in $\omega$ too
small to be observed. Because of the dual ``either/or'' nature of these bounds, it
is impossible with our methods to bound any the $c_{(jk)}$ coefficients
individually.
Yet we can learn something about the characteristic order of
magnitude of the coefficients, assuming there are no special cancellations.
However, there may well be special cancellations; we know that if $c_{(jk)}$ is
cylindrically symmetric---certainly not an unreasonable possibility---there will be
no signal. So we must be careful making generalizations.

So it is important to understand which components of $c$ these kinds of measurements
are actually sensitive to.
The measurable quantities are really the eigenvalue
differences, but all three differences must be nonzero in order for the Lorentz
violation to give a nonzero signal. The eigenvalue differences are
bounded in two distinct
ways, but in general, the relevant sensitivities will be tied together
experimentally, and we expect $\varpi$ and
$\Delta\omega$ not to be too different.
At 90\% confidence (this confidence level reflecting the probability that the
wobbles are not suppressed by a too small $\alpha<\frac{1}{3}$), we may place the
following bound on the $c_{jk}$ coefficients:
\begin{equation}
\label{eq-combbound}
\min\left(\left|c_{11}-c_{22}\right|,\left|c_{11}-c_{33}\right|,\left|c_{22}-c_{33}
\right|\right)<\left[\frac{\max\left(\varpi/2,36\Delta\omega\right)}{\omega_{0}}
\right].
\end{equation}
This is still in the frame where $c_{(jk)}$ is diagonal, so $c_{11}$, $c_{22}$, and
$c_{33}$ are eigenvalues. The bound (\ref{eq-combbound}) is therefore
effectively independent of the pulsar orientation. This is actually advantageous,
because many pulsars have largely unknown spin orientations.

To get actual bounds on the $c$ coefficients, we must now look at data from real
pulsars. We shall rely primarily on data from the pulsar PSR B1509--58. This
neutron star has been under observation for decades, with no glitches
(small but abrupt changes in the period) observed.
The timing data have been analyzed very carefully, in a fully phase-coherent
fashion~\cite{ref-livingstone1}. The pulsar has a frequency $\nu$ of about 6.6 Hz,
known to a fractional accuracy of about $5\times 10^{-10}$. The highest-frequency
residuals seen in the timing plots in~\cite{ref-livingstone1} have periods on the
order of 2 years.
(The dominant residuals have periods in the 5--10 year range.)
The precision with which the frequency is known controls the strength of our limit,
and we have that
\begin{equation}
\label{eq-psrbound}
\min\left(\left|c_{11}-c_{22}\right|,\left|c_{11}-c_{33}\right|,\left|c_{22}-c_{33}
\right|\right)<1.7\times 10^{-8}
\end{equation}
at 90\% confidence. The limit that could be derived from $\varpi$ would be an order
of magnitude better, if the resolution of the instantaneous frequency
could be improved.

As stated, this bound initially appears to be at 90\% confidence. What limits this is
the possibility that the pulsar's spin direction could be very closely aligned with
an eigenvector of $c_{(jk)}$, which would suppress the variation in $\omega$.
However, it seems even more unlikely that there is a quirk in the orientation of
PSR B1509--58 when the data from other pulsars are included. For example, the data
from PSR B0540--69 have been subjected to the same kind of
analysis~\cite{ref-livingstone2}. The analysis is trickier, since PSR B0540--69 has
bee observed to glitch, and a much more complicated fitting procedure is required.
Naively, the results seem to imply an even stronger bound than (\ref{eq-psrbound}),
although the difficulties of the data analysis leave some lingering questions.
We therefore treat this data as merely further confirmation that the result
(\ref{eq-psrbound}) is not merely the result of statistical fluke.

It also is not completely impossible, although it seems very unlikely, that some
other
kind of effect could be masking a signal for Lorentz violation. For example, orbital
motions at a frequency close to $2\Omega$ would produce Doppler shifts affecting the
observed pulsation
frequency, and there could conceivably be cancellation between the two
effects. However, this possibility requires a significant conspiracy of
circumstances. In fact, the bound (\ref{eq-psrbound}) is probably still relatively
conservative.

The bound (\ref{eq-psrbound}) is complementary to the bounds found
in~\cite{ref-kost6}. Clean bounds on just $c$ have previously only been available
for $c_{XX}-c_{XY}$ and $c_{XY}$. These are different degrees of
freedom than are constrained here. The limits in~\cite{ref-kost6} are
very tight, at the $10^{-27}$ level, much better than pulsar bounds.
However, there are lingering questions about their validity, because they rely
on reanalyses of older experimental data sets.
So the present astrophysical bounds are interesting because they are more secure and
because of their complementary structure.

\section*{Acknowledgments}
The author is grateful to V. A. Kosteleck\'{y}, E. Pfister-Altschul, Q. Bailey,
J. Tasson, and M. Berger for helpful discussions.
This work is supported in part by funds provided by the U. S.
Department of Energy (D.O.E.) under cooperative research agreement
DE-FG02-91ER40661.


\begin{thebibliography}{99}

\bibitem{ref-kost18}V. A. Kosteleck\'{y}, S. Samuel, Phys. Rev. D {\bf 39}, 683
(1989).
\bibitem{ref-bluhm1}R. Bluhm, V. A. Kosteleck\'{y}, N. Russell, Phys. Rev.
Lett. {\bf 79}, 1432 (1997).
\bibitem{ref-bluhm2}R. Bluhm, V. A. Kosteleck\'{y}, N. Russell, Phys. Rev. D
{\bf 57}, 3932 (1998).
\bibitem{ref-gabirelse}G. Gabrielse, A. Khabbaz, D. S. Hall, C. Heimann, H.
Kalinowsky, W. Jhe, Phys. Rev. Lett. {\bf 82}, 3198 (1999).
\bibitem{ref-dehmelt1}H. Dehmelt, R. Mittleman, R. S. Van Dyck, Jr., P.
Schwinberg, Phys. Rev. Lett. {\bf 83}, 4694 (1999).
\bibitem{ref-bluhm3}R. Bluhm, V. A. Kosteleck\'{y}, N. Russell , Phys. Rev.
Lett. {\bf 82}, 2254 (1999).
\bibitem{ref-phillips}D. F. Phillips, M. A. Humphrey, E. M. Mattison, R. E.
Stoner, R. F. C. Vessot, R. L. Walsworth , Phys. Rev. D {\bf 63}, 111101 (R)
(2001).
\bibitem{ref-kost8}R. Bluhm, V. A. Kosteleck\'{y}, C. D. Lane, Phys. Rev. Lett.
{\bf 84}, 1098 (2000).
\bibitem{ref-hughes}V. W. Hughes, {\em et al.}, Phys. Rev. Lett. {\bf 87},
111804 (2001).
\bibitem{ref-kost9}R. Bluhm, V. A. Kosteleck\'{y}, Phys. Rev. Lett. {\bf 84},
1381 (2000).
\bibitem{ref-heckel2}B. R. Heckel, C. E. Cramer, T. S. Cook, S. Schlamminger, E. G.
Adelberger, U. Schmidt, Phys. Rev. Lett. {\bf 97}, 021603 (2006).
\bibitem{ref-berglund}C. J. Berglund, L. R. Hunter, D. Krause, Jr., E. O.
Prigge, M. S. Ronfeldt, S. K. Lamoreaux, Phys. Rev. Lett. {\bf 75}, 1879 (1995).
\bibitem{ref-kost6}V. A. Kosteleck\'{y}, C. D. Lane, Phys. Rev. D {\bf 60},
116010 (1999).
\bibitem{ref-bear}D. Bear, R. E. Stoner, R. L. Walsworth, V. A. Kosteleck\'{y},
C. D. Lane, Phys. Rev. Lett. {\bf 85}, 5038 (2000).
\bibitem{ref-wolf}P. Wolf, F. Chapelet, S. Bize, A. Clairon,  Phys. Rev. Lett.
{\bf 96}, 060801 (2006).
\bibitem{ref-muller1}H. M\"{u}ller, S. Herrmann, C. Braxmaier, S. Schiller, A.
Peters, Phys. Rev. Lett. {\bf 91}, 020401 (2003).
\bibitem{ref-stanwix}P. L. Stanwix, M. E. Tobar, P. Wolf, M. Susli, C. R. Locke,
E. N. Ivanov, J. Winterflood, F. van Kann, Phys. Rev. Lett. {\bf 95}, 040404
(2005).
\bibitem{ref-saathoff}G. Saathoff, S. Karpuk, U. Eisenbarth, G. Huber, S. Krohn, R.
Mu\~{n}oz Horta, S. Reinhardt, D. Schwalm, A. Wolf, G. Gwinner, Phys. Rev. Lett.
{\bf 91}, 190403 (2003).
\bibitem{ref-lane1}C. D. Lane, Phys. Rev. D {\bf 72}, 016005 (2005).
\bibitem{ref-kost10}V. A. Kosteleck\'{y}, Phys. Rev. Lett. {\bf 80}, 1818
(1998).
\bibitem{ref-kost7}V. A. Kosteleck\'{y}, Phys. Rev. D {\bf 61}, 016002 (2000).
\bibitem{ref-hsiung}Y. B. Hsiung, Nucl. Phys. Proc. Suppl. {\bf 86}, 312
(2000).
\bibitem{ref-abe}K. Abe {\em et al.}, Phys. Rev. Lett. {\bf 86}, 3228 (2001).
\bibitem{ref-carroll1}S. M. Carroll, G. B. Field, R. Jackiw, Phys. Rev. D
{\bf 41}, 1231 (1990).
\bibitem{ref-carroll2}S. M. Carroll, G. B. Field, Phys. Rev. Lett. {\bf 79},
2394 (1997).
\bibitem{ref-kost11}V. A. Kosteleck\'{y}, M. Mewes, Phys. Rev. Lett. {\bf 87},
251304 (2001).
\bibitem{ref-kost21}V. A. Kosteleck\'{y}, M. Mewes, hep-ph/0607084.
\bibitem{ref-jacobson1}T. Jacobson, S. Liberati, D. Mattingly, Nature {\bf 424},
1019 (2003).
\bibitem{ref-altschul6}B. Altschul, Phys. Rev. Lett. {\bf 96}, 201101 (2006).
\bibitem{ref-kost1}D. Colladay, V. A. Kosteleck\'{y}, Phys. Rev. D {\bf 55},
6760 (1997).
\bibitem{ref-kost2}D. Colladay, V. A. Kosteleck\'{y}, Phys. Rev. D {\bf 58},
116002 (1998).
\bibitem{ref-kost12}V. A. Kosteleck\'{y}, Phys. Rev. D {\bf 69}, 105009 (2004).
\bibitem{ref-kost3}V. A. Kosteleck\'{y}, R. Lehnert, Phys. Rev. D {\bf 63},
065008 (2001).
\bibitem{ref-kost4}V. A. Kosteleck\'{y}, C. D. Lane, A. G. M. Pickering,
Phys. Rev. D {\bf 65}, 056006 (2002).
\bibitem{ref-altschul5}B. Altschul, V. A. Kosteleck\'{y}, Phys. Lett. B {\bf 628},
106 (2005).
\bibitem{ref-nordtvedt}K. Nordtvedt, Astrophys. J. {\bf 320}, 871 (1987).
\bibitem{ref-colladay}D. Colladay, P. McDonald, Phys. Rev. D {\bf 70} 125007 (2004).
\bibitem{ref-livingstone1}M. A. Livingstone, V. M. Kaspi, F. P. Gavril, R. N.
Manchester, Astrophys. J. {\bf 619}, 1046 (2005).
\bibitem{ref-livingstone2}M. A. Livingstone, V. M. Kaspi, F. P. Gavril, Astrophys. J.
{\bf 633}, 1095 (2005).


\end{thebibliography}
\end{document}